\documentclass[twocolumn,final, 5p, sort&compress, twocolumn]{elsarticle}

\usepackage[T1]{fontenc}
\usepackage[latin9]{inputenc}
\usepackage{color}
\usepackage{amsmath}
\usepackage{amssymb}
\usepackage{graphicx}
\usepackage{wasysym}
\usepackage[unicode=true,
 bookmarks=true,bookmarksnumbered=false,bookmarksopen=false,
 breaklinks=false,pdfborder={0 0 1},backref=false,colorlinks=true]
 {hyperref}
\hypersetup{
 urlcolor=blue}
\begin{document}

\begin{frontmatter}{}

\title{The correlational entropy production during the local relaxation
in a many body system with Ising interactions}

\author{Tai Kang }

\author{Sheng-Wen Li}

\ead{lishengwen@bit.edu.cn}

\address{Center for Quantum Technology Research, and Key Laboratory of Advanced
Optoelectronic Quantum Architecture and Measurements, School of Physics,
Beijing Institute of Technology, Beijing 100081, China}
\begin{abstract}
Isolated quantum systems follow the unitary evolution, which guarantees
the full many body state always keeps a constant entropy as its initial
one. In comparison, the local subsystems exhibit relaxation behavior
and evolve towards certain steady states, which is called the local
relaxation. Here we consider the local dynamics of finite many body
system with Ising interaction. In both strong and weak coupling situations,
the local observables exhibit similar relaxation behavior as the macroscopic
thermodynamics; due to the finite size effect, recurrence appears
after a certain typical time. Especially, we find that the total correlation
of this system approximately exhibits a monotonic increasing envelope
in both strong and weak coupling cases, which corresponds to the irreversible
entropy production in the standard macroscopic thermodynamics. Moreover,
the possible maximum of such total correlation  calculated under proper
constraints also coincides well with the exact result of time dependent
evolution.
\end{abstract}

\end{frontmatter}{}

\section{Introduction\label{sec:Introduction}}

A macroscopic thermodynamic system always tends to relax to the thermal
equilibrium state after a long enough time in spite of its initial
state. However, it is also known that an isolated quantum system with
a finite number of degrees of freedom (DoF) follows the unitary evolution.
As a result, the full many body system always keeps a constant entropy
as its initial state, which looks different from the above macroscopic
relaxation behavior, unless certain specific averaging approaches
are taken into consideration, e.g., based on time or random defect
configurations \citep{hobson_irreversibility_1966,prigogine_time_1978,mackey_dynamic_1989,evans_fluctuation_2002,uffink_compendium_2006,swendsen_explaining_2008,landi_irreversible_2021}.

On the other hand, when focusing on the local parts of a many body
system, the local dynamics naturally manifests similar relaxation
behavior as the macroscopic thermodynamics. Although the whole isolated
system always follows the unitary evolution and keeps a constant entropy,
the dynamics of a local site exhibits an oscillating decay behavior,
which seems relaxing towards a certain steady state, thus this is
called the local relaxation \citep{cramer_exact_2008,li_hierarchy_2021,flesch_probing_2008,eisert_quantum_2015}.
For a finite system size, the local relaxation would come across a
``recurrence'' after a typical time: the well ordered oscillatory
decay behavior suddenly appears ``random'' \citep{cramer_exact_2008,li_hierarchy_2021,flesch_probing_2008,eisert_quantum_2015,hanggi_reaction-rate_1990,zwanzig_nonequilibrium_2001}.
With the increase of the system size, the recurrence appears much
later, thus it does not show up in practice. 

But the entropy of each local system cannot indicate the irreversible
entropy production behavior either, since it increases and decreases
from time to time. In comparison, it is found that the total correlation
 in the many body system \citep{watanabe_information_1960,groisman_quantum_2005,zhou_irreducible_2008,goold_total_2015,anza_logarithmic_2020,landi_irreversible_2021}
could manifest the irreversible entropy production similar as the
standard thermodynamics \citep{lebowitz_macroscopic_1993,li_production_2017,you_entropy_2018,li_correlation_2019,anza_logarithmic_2020,li_hierarchy_2021}.
In this paper, we study the dynamical behavior of the total correlation
 during the local relaxation in an $N$-body system with Ising interactions.
In both strong and weak coupling regimes, which correspond to the
two phases of the Ising model with distinct physical properties though,
the system dynamics exhibits quite similar local relaxation and recurrence
behaviors, except in the strong coupling case the system dynamics
contains more violent fluctuations. 

Especially, it turns out the evolution of the total correlation roughly
exhibits a monotonic increasing behavior in both strong and weak coupling
cases, and approaches towards a steady value during the local relaxation
process, which is similar to the irreversible entropy increase in
the standard thermodynamics \citep{lebowitz_macroscopic_1993,li_production_2017,you_entropy_2018,li_correlation_2019}.
Moreover, with the help of the Lagrangian multipliers, we calculate
the possible maximum of the total correlation may achieve under proper
constraints determined by the initial state, and it turns out such
correlation maximization coincides quite well with the exact numerical
result obtained from time dependent evolution. In this sense, the
correlational entropy production well corresponds to the irreversible
entropy production in the standard thermodynamics. We also show that
in most common physical conditions, the correlational entropy production
could reduce to the standard entropy production based on the thermal
entropy $\text{\dj}Q/T$.

The paper is arranged as follows. In Sec.\,2 we study the dynamics
of the Ising model. In Sec.\,3 we analyze the recurrence behavior
in the local dynamics with different coupling strengths. In Sec.\,4
we show the dynamics of the total correlation in this system. In Sec.\,5
we show the correlational entropy production could reduce to the entropy
production in the standard thermodynamics. The conclusion is drawn
in Sec.\,6. 

\section{Dynamics in a many-body system with Ising interaction\label{sec:Dynamics-in-the}}

Here we consider an isolated many body system, which contains $N$
two-level systems (TLSs) with the periodic boundary condition. The
$N$ TLSs interact with the near neighbors via the Ising interaction,
which is described by the Hamiltonian \citep{sachdev_quantum_2011}:
\begin{equation}
\hat{H}=\sum_{n=0}^{N-1}\frac{\omega}{2}\,\hat{\sigma}_{n}^{z}+J\,\hat{\sigma}_{n}^{x}\hat{\sigma}_{n+1}^{x}:=\hat{H}_{0}+\hat{V}_{J}.\label{eq:H-0}
\end{equation}
Here $\hat{\sigma}_{n}^{x,z}$ are two Pauli matrices, and $\hat{\sigma}_{n}^{z}:=|\mathrm{e}\rangle_{n}\langle\mathrm{e}|-|\mathrm{g}\rangle_{n}\langle\mathrm{g}|$,
$\hat{\sigma}_{n}^{x}:=\hat{\sigma}_{n}^{+}+\hat{\sigma}_{n}^{-}$,
with $\hat{\sigma}_{n}^{+}:=(\hat{\sigma}_{n}^{-})^{\dagger}=|\mathrm{e}\rangle_{n}\langle\mathrm{g}|$.
Here $|\mathrm{e}\rangle_{n}$, $|\mathrm{g}\rangle_{n}$ are the
excited and ground states of the $n$-th TLS. The on-site energies
($\omega\ge0$) of all the $N$ sites are equal, and $J$ is the interaction
strength.

By applying the Jordan-Wigner transform \citep{sachdev_quantum_2011},
\begin{align}
\hat{\sigma}_{n}^{z} & =2\hat{d}_{n}^{\dagger}\hat{d}_{n}-1,\qquad\hat{\sigma}_{n}^{+}=\hat{d}_{n}^{\dagger}\,\prod_{i=0}^{n-1}\left(-\hat{\sigma}_{i}^{z}\right),\nonumber \\
\hat{\sigma}_{n}^{x} & =\left(\hat{d}_{n}+\hat{d}_{n}^{\dagger}\right)\,\prod_{i=0}^{n-1}\left(-\hat{\sigma}_{i}^{z}\right),\label{eq:JW}
\end{align}
 the above Hamiltonian becomes a fermionic one, $\hat{H}=\sum_{n}\,\omega\,\hat{d}_{n}^{\dagger}\hat{d}_{n}+J(\hat{d}_{n}^{\dagger}\hat{d}_{n+1}+\hat{d}_{n}^{\dagger}\hat{d}_{n+1}^{\dagger}+\text{h.c.})$.
Further, under the Fourier transform\footnote{Without loss of generality, we assume $N$ is odd. Generally speaking,
in the thermodynamic limit $N\rightarrow\infty$, the thermodynamic
properties of the system do not depend on the parity of $N$.} $\hat{d}_{n}=\sum_{k}\,\hat{c}_{k}\,e^{-i\frac{2\pi}{N}kn}/\sqrt{N}$,
with $k=-\frac{1}{2}(N-1),\ldots,0\ldots,\frac{1}{2}(N-1)$, the Hamiltonian
(\ref{eq:H-0}) becomes $\hat{H}=\frac{1}{2}\sum\mathbf{c}_{k}^{\dagger}\cdot\mathbf{H}_{k}\cdot\mathbf{c}_{k}$,
where $\mathbf{c}_{k}:=(\hat{c}_{k},\,\hat{c}_{-k}^{\dagger})^{T}$
and 
\begin{equation}
\mathbf{H}_{k}=\left[\begin{array}{cc}
\frac{1}{2}\omega+2J\cos\frac{2\pi}{N}k & i2J\sin\frac{2\pi}{N}k\\
-i2J\sin\frac{2\pi}{N}k & -\left(\frac{1}{2}\omega+2J\cos\frac{2\pi}{N}k\right)
\end{array}\right].
\end{equation}
Then the system Hamiltonian can be diagonalized as $\hat{H}=\sum_{k>0}\,\varepsilon_{k}\,\hat{\gamma}_{k}^{\dagger}\hat{\gamma}_{k}-\varepsilon_{k}\,\hat{\gamma}_{-k}\hat{\gamma}_{-k}^{\dagger}$,
where $\hat{\gamma}_{k}:=\cos\theta_{k}\,\hat{c}_{k}+i\sin\theta_{k}\,\hat{c}_{-k}^{\dagger}$
is the Bogoliubov operator, and $\varepsilon_{k}$ is the eigen mode
energy \citep{sachdev_quantum_2011} 
\begin{gather}
\varepsilon_{k}=\big[(\frac{1}{2}\omega+2J\cos\frac{2\pi}{N}k)^{2}+4J^{2}\sin^{2}\frac{2\pi}{N}k\big]^{\frac{1}{2}},\nonumber \\
\tan2\theta_{k}=\frac{2J\sin\frac{2\pi}{N}k}{\frac{1}{2}\omega+2J\cos\frac{2\pi}{N}k}.\label{eq:BT}
\end{gather}

Based on these results, now we could obtain the exact time-dependent
evolution of this isolated system. We consider that the initial state
of the system is $|\varPsi_{0}\rangle=|\mathrm{e},\mathrm{g},\mathrm{g},\ldots\rangle$,
namely, site-0 starts from the excited state $|\mathrm{e}\rangle_{0}$,
and all the other sites start from the ground state $|\mathrm{g}\rangle_{n}$.
Here site-0 could be regarded as an open ``system'', while all the
other $(N-1)$ TLSs build up a finite ``bath''.

By using the Jordan-Wigner transform (\ref{eq:JW}), such an initial
state also can be rewritten as $|\varPsi_{0}\rangle=\hat{d}_{n=0}^{\dagger}\,|\mathbf{0}\rangle$,
where $|\mathbf{0}\rangle$ is the vacuum state of the fermion system.
Then we obtain the system state $|\varPsi_{t}\rangle$ at an arbitrary
time, which gives 
\begin{align}
|\varPsi_{t}\rangle & =e^{-i\hat{H}t}\,\hat{d}_{0}^{\dagger}|\mathbf{0}\rangle=\sum_{n=0}^{N-1}\Phi_{n}^{(N)}(t)\ \hat{d}_{n}^{\dagger}|\mathbf{0}\rangle,\label{eq:Psi-t}\\
\Phi_{n}^{(N)}(t) & :=\sum_{k}\frac{e^{-i\frac{2\pi}{N}kn}}{N}\left(\cos^{2}\theta_{k}e^{i\varepsilon_{k}t}+\sin^{2}\theta_{k}e^{-i\varepsilon_{k}t}\right).\nonumber 
\end{align}
Here we call $\Phi_{n}^{(N)}(t)$ as a coherence function. 

In principle now all the observable expectations can be calculated
from $|\varPsi_{t}\rangle$. With the help of $\Phi_{n}^{(N)}(t)$
and the Jordan-Wigner transform (\ref{eq:JW}), the excitation probability
$p_{n,\mathrm{e}}(t):=\langle\mathrm{e}|\rho_{n}(t)|\mathrm{e}\rangle_{n}$
of the reduced density state $\rho_{n}$ of each site-$n$ gives 
\begin{equation}
p_{n,\mathrm{e}}(t)=\langle\varPsi_{t}|\hat{d}_{n}^{\dagger}\hat{d}_{n}|\varPsi_{t}\rangle=|\Phi_{n}^{(N)}(t)|^{2},\label{eq:pe}
\end{equation}
and the total excitation of all the $N$ sites is $\langle\hat{\text{\textsc{n}}}_{\mathrm{t}}\rangle=\sum_{n}\,|\Phi_{n}^{(N)}(t)|^{2}$.
Besides, from Eqs.\,(\ref{eq:JW}, \ref{eq:Psi-t}) it can be verified
that the non-diagonal terms of the density state $_{n}\langle\mathrm{e}|\rho_{n}(t)|\mathrm{g}\rangle_{n}$
of each site always keep zero during the evolution \citep{wu_exact_2021}.

Notice that, when the coupling strength is quite weak ($J\ll\omega$),
in the interaction picture of $\hat{H}_{0}=\frac{1}{2}\omega\sum\hat{\sigma}_{n}^{z}$,
the interaction in the Hamiltonian (\ref{eq:H-0}) becomes 
\begin{align}
\hat{V}_{J}(t) & =J\sum_{n}\left(\hat{\sigma}_{n}^{+}e^{i\omega t}+\hat{\sigma}_{n}^{-}e^{-i\omega t}\right)\left(\hat{\sigma}_{n+1}^{+}e^{i\omega t}+\hat{\sigma}_{n+1}^{-}e^{-i\omega t}\right)\nonumber \\
 & \simeq J\sum_{n}\left(\hat{\sigma}_{n}^{+}\hat{\sigma}_{n+1}^{-}+\hat{\sigma}_{n}^{-}\hat{\sigma}_{n+1}^{+}\right),
\end{align}
where the oscillating terms with frequencies $2\omega$ are neglected
due to the rotating-wave approximation (RWA). Returning back to the
Schr\"odinger picture the Hamiltonian becomes 
\begin{equation}
\hat{H}=\sum_{n=0}^{N-1}\frac{\omega}{2}\,\hat{\sigma}_{n}^{z}+J(\hat{\sigma}_{n}^{+}\hat{\sigma}_{n+1}^{-}+\hat{\sigma}_{n}^{-}\hat{\sigma}_{n+1}^{+}).\label{eq:XX}
\end{equation}
This is also known as the quantum \emph{XX} model \citep{sachdev_quantum_2011},
which guarantees the total excitation ($\hat{\text{\textsc{n}}}_{\mathrm{t}}\sim\sum\hat{\sigma}_{n}^{z}$)
is conserved, while the original Ising model does not. Therefore,
the behavior of the system dynamics would be similar as the quantum
\emph{XX} model in the weak coupling regime \citep{li_hierarchy_2021},
while they would exhibit significant differences when $J$ is strong.
Such a comparison could be well seen in our numerical results below.

\section{The recurrence in the local dynamics\label{sec:The-recurrence-behavior}}

Now the full evolution of the $N$-body state is obtained exactly
{[}Eq.\,(\ref{eq:Psi-t}){]}. Since the whole isolated system follows
the unitary evolution, if we do not consider the average over time
or random defect configurations, the full $N$-body state $|\varPsi_{t}\rangle$
would always keep a pure state during the evolution. Namely, indeed
$|\varPsi_{t}\rangle$ is never approaching any canonical thermal
state as $\rho_{\text{th}}\sim\exp(-\hat{H}/k_{\text{\textsc{b}}}T)$. 

\begin{figure}[t]
\includegraphics[width=1\columnwidth]{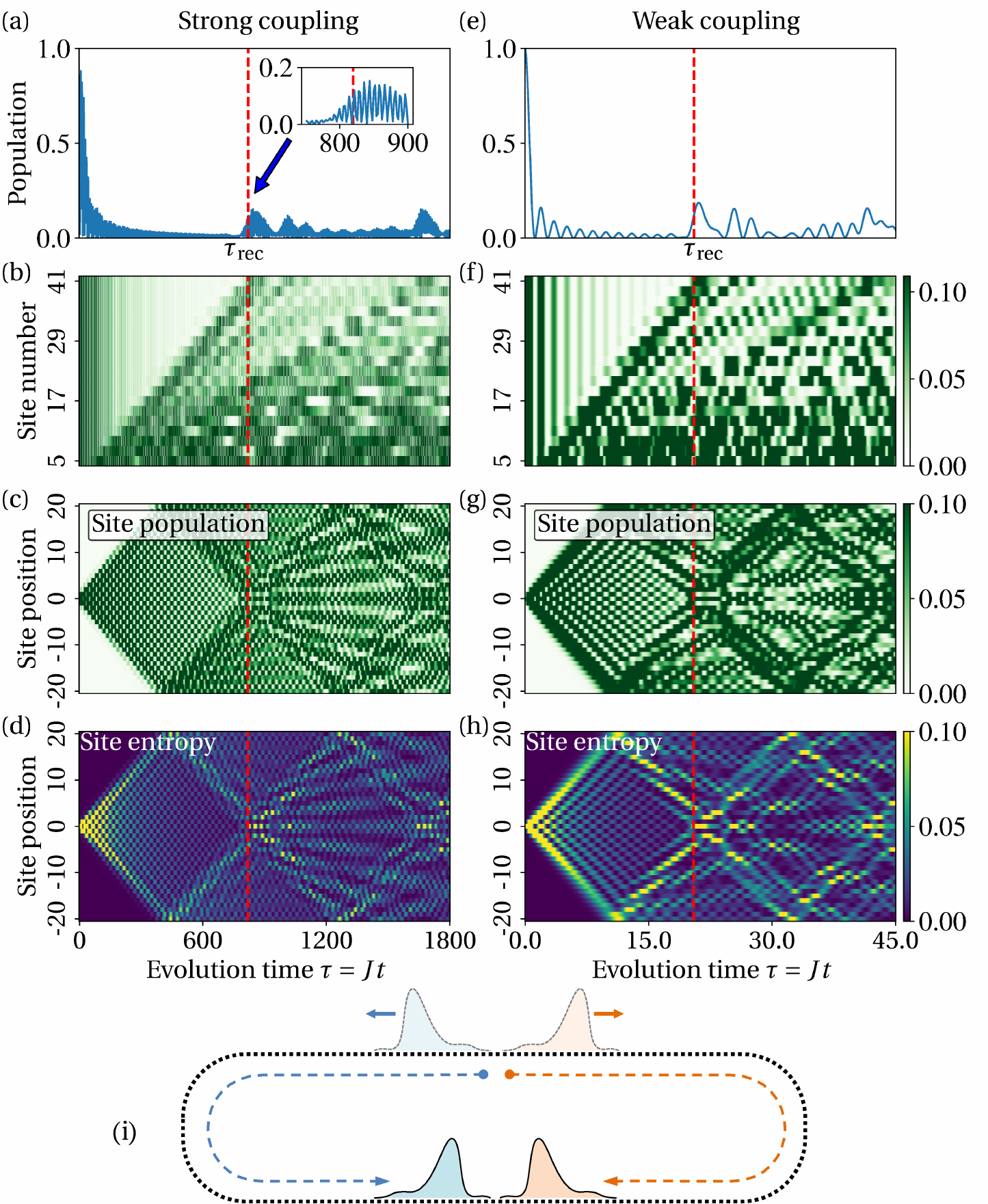}

\caption{(a, e) The population evolution of site-$0$ $p_{0,\mathrm{e}}(t)=|\Phi_{0}^{(N=41)}(t)|^{2}$
in the strong ($\omega=0.1J$) and weak ($\omega=40J$) coupling regimes.
(b, f) The scaling behavior of $p_{0,\mathrm{e}}(t)=|\Phi_{0}^{(N)}(t)|^{2}$
for different site numbers $N$. (c, g) The evolution of the local
excitations $p_{n,\mathrm{e}}(t)$ of all the $N$ sites. (d, h) The
evolution of the von Neumann entropy $S[\rho_{n}(t)]$ of all the
$N$ sites. All the site numbers are set as $N=41$ except (b, f).
(i) Demonstration for the propagation of the local excitations along
the chain. Throughout this paper, we set $J=1$ as the energy unit,
and all the evolutions are measured by the unitless time $\tau=Jt$.}

\label{fig:kktt}
\end{figure}

In contrast, if we focus on the states of each local site, we would
see the local dynamics exhibits similar behavior as the relaxation
in macroscopic thermodynamics \citep{zwanzig_nonequilibrium_2001,cramer_exact_2008,li_hierarchy_2021}.
In Fig.\,\ref{fig:kktt}(a, d), we show the evolution of the excitation
probability $p_{n=0,\mathrm{e}}(t)$ of site-0 in both the strong
and weak coupling regimes. Before a certain typical time $t\apprle t_{\text{rec}}$
(the vertical dashed red lines in Fig.\,\ref{fig:kktt}), they both
exhibits an oscillating decay behavior, which is quite similar as
the relaxation in macroscopic thermodynamics. Namely, within the observation
time $t<t_{\text{rec}}$, the population of site-0 (the ``system'')
seems relaxing towards $p_{0,\mathrm{e}}\rightarrow0$ as its steady
state. 

But the decaying behavior suddenly changes around $t\sim t_{\text{rec}}$,
the excitation probability $p_{0,\mathrm{e}}(t)$ shows a sudden increase
and then looks ``random''. This was referred as a ``recurrence''
due to the finite size effect \citep{cramer_exact_2008,zwanzig_nonequilibrium_2001}.
With the increase of the system size $N$, the recurrence time $t_{\text{rec}}$
also increases linearly {[}Fig.\,\ref{fig:kktt}(b, f){]}. When $N\rightarrow\infty$,
the recurrence time would also be postponed and approach infinity.
Thus at a finite time $t\ll t_{\text{rec}}$, only the irreversible
relaxation behavior can be observed in practice \citep{cramer_exact_2008,zwanzig_nonequilibrium_2001}.

Besides, in the relaxation region before the recurrence $t\apprle t_{\text{rec}}$,
notice that the decaying profiles for different system size $N$ are
almost same with each other {[}Fig.\,\ref{fig:kktt}(b, f){]}. That
indicates, once  $J,\,\omega$ are fixed, the relaxation speed of
this system is also fixed, which does not depend on the system size
$N$. For a larger system size $N$, the full diffusion over the whole
many-body system would cost more time.

It is worth noting that, such recurrences appear in both strong and
weak coupling regimes (Fig.\,\ref{fig:kktt}). In the weak coupling
regime $J\ll\omega$, the system does show a dynamics similar as the
\emph{XX} model (\ref{eq:XX}) as mentioned above \citep{cramer_exact_2008,li_hierarchy_2021,zwanzig_nonequilibrium_2001}.
In the strong coupling regime, the system dynamics also exhibits local
relaxation and recurrence behaviors as the weak coupling situation,
but contains more high frequency oscillations. For different parameter
settings in both weak and strong regimes, we find that the recurrence
time can be evaluated as $t_{\text{rec}}\equiv N(J^{-2}+16\omega^{-2})^{\frac{1}{2}}/2$,
which is estimated from $t_{\text{rec}}\sim2\pi/\delta\varepsilon$,
with $\delta\varepsilon$ as the largest gap of the mode energies
$\{\varepsilon_{k}\}$ (appearing around $2\pi k/N\sim\pm\pi/2$).

In Fig.\,\ref{fig:kktt}(c, g) we show the evolution of the excitation
probabilities $p_{n,\mathrm{e}}(t)$ of all the $N$ sites. In both
strong and weak coupling cases, we could clearly see a propagation
process of the site excitations along the chain. Starting from site-0,
the local excitations propagate along the two sides of the chain.
For a finite system size, such propagations would meet each other
at the periodic boundaries at $n\sim\pm N/2$, and then regathers
back to site-0 again. This is just why the system exhibits its recurrence
around $t\simeq t_{\text{rec}}$ {[}see the demonstrations in Fig.\,\ref{fig:kktt}(i){]}.
After the recurrence, the system dynamics is combined with the regathered
propagations together, thus appears more random. This propagation
picture well explains why the recurrences appear in both the strong
and weak coupling cases.

Indeed such propagation and regathering behaviors happen again and
again, thus similar recurrence behavior could also appear around $t\simeq qt_{\text{rec}}$,
with $q=1,2,3\ldots$ The system dynamics appears more and more ``random''
after each recurrence, and thus we call them hierarchy recurrences
\citep{li_hierarchy_2021}. 

In Fig.\,\ref{fig:kktt}(d, h) we show the evolution of the von Neumann
entropy of each site, i.e., $S[\rho_{n}]\equiv-\mathrm{tr}[\rho_{n}\ln\rho_{n}]$.
The entropy of each site increases and decreases from time to time,
and also exhibit a propagation pattern along the chain in both both
the strong and weak coupling situations. From this picture, we could
see as long as there exist similar propagations of the local excitations,
the local relaxations and such recurrences could appear in spite of
the interaction strength and type.

\section{Production of the total correlation \label{sec:Total-correlation-entropy}}

Now we see the local observables well exhibit the relaxation behaviors
similar as the macroscopic thermodynamics, but the entropy of each
site increases and decreases from time to time, which is different
from the irreversible entropy production behavior in macroscopic thermodynamics.
On the other hand, as an isolated system, the entropy of the whole
system state always keeps the same as its initial state due to the
the unitary evolution. Indeed, since the full $N$-body state here
is not a canonical equilibrium state, the standard entropy production
in macroscopic thermodynamics based on thermal entropy $dS=\text{\dj}Q/T$
cannot be applied here.

Notice that, in practical observations, the full $N$-body state is
usually not directly assessable for local measurements, and it is
the few-body observables that are directly measured \citep{swendsen_explaining_2008,li_correlation_2019,strasberg_entropy_2019}.
Thus, here we consider the \emph{total correlation} in this system,
which is defined by \citep{watanabe_information_1960,groisman_quantum_2005,zhou_irreducible_2008,goold_total_2015,anza_logarithmic_2020,landi_irreversible_2021}
\begin{equation}
\mathbf{C}[\hat{\boldsymbol{\rho}}]:=\sum_{\alpha=0}^{N-1}S[\hat{\rho}_{\alpha}]-S[\hat{\boldsymbol{\rho}}].\label{eq:C=00005Brho=00005D}
\end{equation}
Here $S[\hat{\boldsymbol{\rho}}]$ is the von Neumann entropy of the
full $N$-body state $\hat{\boldsymbol{\rho}}$, which does not change
during the unitary evolution, and $\hat{\rho}_{\alpha}$ are the reduced
one-body states. This total correlation was firstly introduced for
classical systems based on the collective probability distributions
and the Shannon entropy \citep{watanabe_information_1960}, and the
generalization for quantum systems by using density states and the
von Neumann entropy is straightforward \citep{groisman_quantum_2005,zhou_irreducible_2008,goold_total_2015}.
$\mathbf{C}[\hat{\boldsymbol{\rho}}]$ measures the total amount of
all the correlations inside the $N$-body state $\hat{\boldsymbol{\rho}}$
\citep{watanabe_information_1960,groisman_quantum_2005,zhou_irreducible_2008,goold_total_2015,anza_logarithmic_2020,landi_irreversible_2021}.
For $N=2$, the total correlation (\ref{eq:C=00005Brho=00005D}) just
returns the mutual information in a two-body system, which measures
the amount of the bipartite correlation.

\begin{figure}
\includegraphics[width=1\columnwidth]{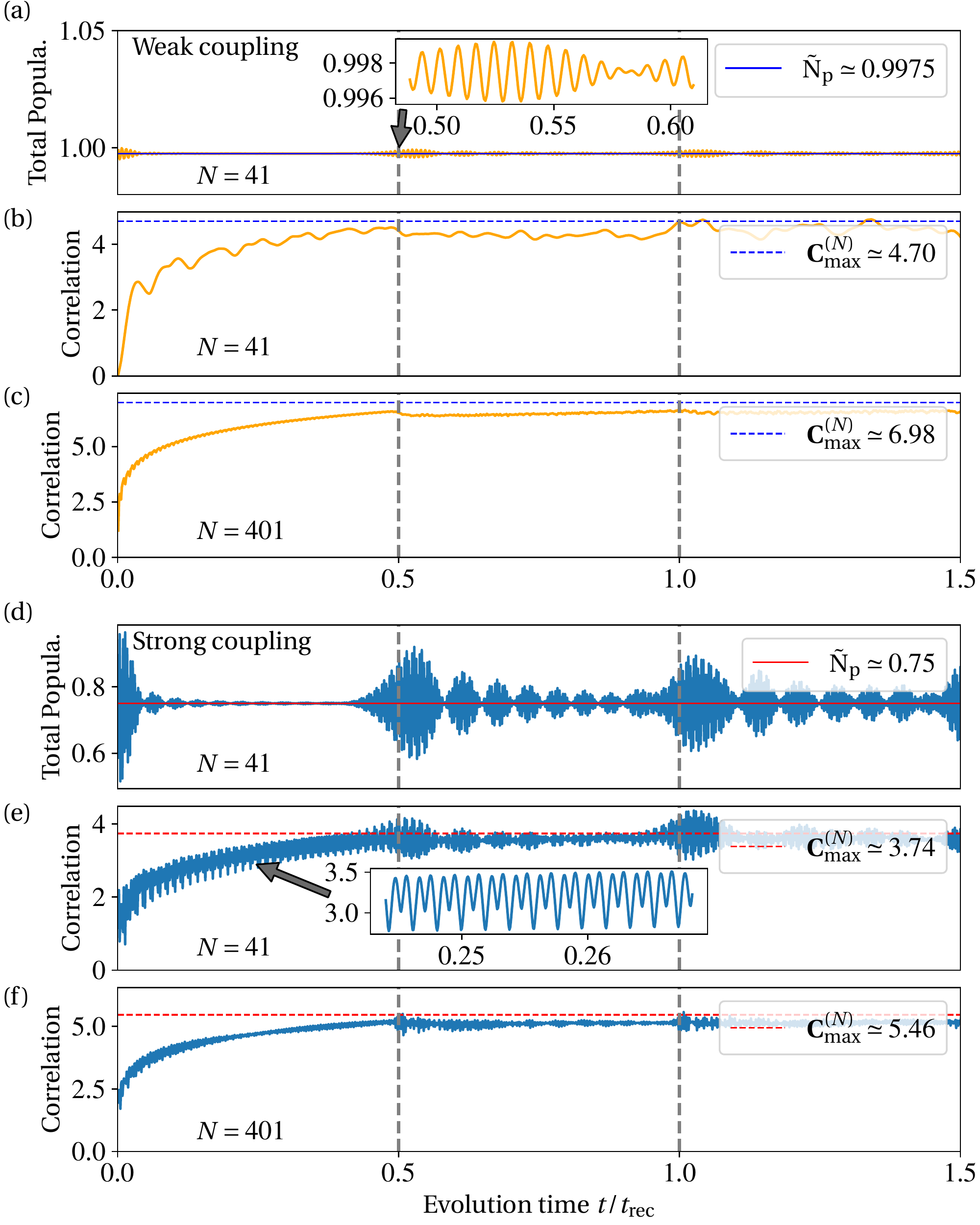}

\caption{The evolution of the total excitation $\langle\hat{\text{\textsc{n}}}_{\mathrm{t}}\rangle$
(a, d), and the total correlation (b, c, e, f) in the weak (a-c, $\omega=40J$)
and strong (d-f, $\omega=0.1J$) coupling regimes. The site numbers
are set as $N=41$ in (a, b, d, e), while $N=401$ in (c, f). The
dashed horizontal lines in (b, c, e, f) are the obtained by the total
correlation maximization (\ref{eq:C}). The solid horizontal lines
in (a, d) are $\tilde{\text{\textsc{n}}}_{\text{\textsc{p}}}\simeq0.9975$
and $\tilde{\text{\textsc{n}}}_{\text{\textsc{p}}}\simeq0.75$ respectively,
which are the total excitations calculated by Eq.\,(\ref{eq:np}).}

\label{fig:TCE}
\end{figure}

In Fig.\,\ref{fig:TCE}(b, c, e, f), we show the time dependent evolution
of the total correlation in both weak and strong coupling regimes,
and they both exhibit an increasing behavior towards a steady value.
In the weak coupling case {[}Fig.\,\ref{fig:TCE}(b, c){]}, the behavior
of the total correlation is similar as the \emph{XX} model \citep{li_hierarchy_2021}.
In the strong coupling case {[}Fig.\,\ref{fig:TCE}(e, f){]}, the
total correlation also exhibits an increasing envelope, but contains
more violent oscillations. Moreover, with the increase of the system
size $N$, the increasing curve in both cases become more and more
smooth {[}Fig.\,\ref{fig:TCE}(c, f){]}. These behaviors are quite
similar with the irreversible entropy production in the standard macroscopic
thermodynamics.

In this sense, we further consider where is the destination that the
total correlation is increasing towards. The possible maximum that
$\mathbf{C}[\hat{\boldsymbol{\rho}}(t)]$ may achieve can be obtained
by the variation approach under proper constraints \citep{jaynes_information_1957}. 

Unlike the quantum \emph{XX} model (\ref{eq:XX}), here the total
excitation of all the $N$ site is not conserved and changes with
time. From the transformations (\ref{eq:JW}, \ref{eq:BT}), the total
excitation $\langle\hat{\text{\textsc{n}}}_{\mathrm{t}}\rangle=\sum_{n}\langle\hat{d}_{n}^{\dagger}(t)\hat{d}_{n}(t)\rangle$
gives 
\begin{align}
\langle\hat{\text{\textsc{n}}}_{\mathrm{t}}\rangle= & \frac{1}{N}\sum_{k}\cos^{4}\theta_{k}-\left(i\sin\theta_{k}\cos\theta_{k}\right)^{2}e^{2i\varepsilon_{k}t}\nonumber \\
 & -\left(i\sin\theta_{k}\cos\theta_{k}\right)^{2}e^{-2i\varepsilon_{k}t}+\sin^{4}\theta_{k}.\label{eq:Nt}
\end{align}

It turns out, in the weak coupling case {[}Fig.\,\ref{fig:TCE}(a){]},
the total excitation exhibits quite small fluctuations with time around
a certain constant, which is consistent with the above discussions
about the reduction to the \emph{XX} model by RWA. Thus, approximately
the total excitation still could be treated as a variation constraint.

When the Ising interaction is strong {[}Fig.\,\ref{fig:TCE}(d){]},
the total excitation exhibits significant fluctuations with time,
which is quite different from the \emph{XX} model. But it is also
worth noting that, especially within the relaxation time $t\apprle t_{\text{rec}}/2$,
the total excitation $\langle\hat{\text{\textsc{n}}}_{\mathrm{t}}\rangle$
seems converging towards a steady value, until a recurrence happens
because of the finite size effect. Indeed the moment that the total
excitation comes across its first recurrence ($\sim t_{\text{rec}}/2$)
is just when the two-side propagations meet each other at the periodic
boundaries at $n\sim\pm N/2$ {[}see Fig.\,\ref{fig:kktt}(i){]}.
In the thermodynamic limit $N\rightarrow\infty$, recurrences do not
appear, and the total excitation $\langle\hat{\text{\textsc{n}}}_{\mathrm{t}}\rangle$
is just relaxing to this steady value irreversibly.

Therefore, such a converging value of the total excitation $\langle\hat{\text{\textsc{n}}}_{\mathrm{t}}\rangle$
could be treated as an approximated constraint when we estimate the
possible maximum where the total correlation is increasing towards.
Further, since $\langle\hat{\text{\textsc{n}}}_{\mathrm{t}}\rangle$
keeps oscillating around a central value, this converging value can
be estimated by the long time average of Eq.\,(\ref{eq:Nt}), which
eliminates all the rotating terms, i.e., \citep{srednicki_chaos_1994,rigol_alternatives_2012,polkovnikov_colloquium:_2011}
\begin{align}
\tilde{\text{\textsc{n}}}_{\text{\textsc{p}}} & :=\lim_{T\rightarrow\infty}\frac{1}{T}\int_{0}^{T}dt\,\langle\hat{\text{\textsc{n}}}_{\mathrm{t}}\rangle=\frac{1}{N}\sum_{k}\cos^{4}\theta_{k}+\sin^{4}\theta_{k}\nonumber \\
= & \frac{1}{2}+\frac{1}{2N}\sum_{k}\frac{(\frac{1}{2}\omega+2J\cos\frac{2\pi}{N}k)^{2}}{(\frac{1}{2}\omega+2J\cos\frac{2\pi}{N}k)^{2}+4J^{2}\sin^{2}\frac{2\pi}{N}k}.\label{eq:np}
\end{align}
 In Fig.\,\ref{fig:TCE}(a, d) we see $\tilde{\text{\textsc{n}}}_{\text{\textsc{p}}}$
does give the central value that $\langle\hat{\text{\textsc{n}}}_{\mathrm{t}}\rangle$
oscillates around {[}the horizontal solid lines{]}. 

In the thermodynamics limit $N\rightarrow\infty$, this central value
becomes ($\tilde{J}\equiv J/\omega$)
\begin{align}
\tilde{\text{\textsc{n}}}_{\text{\textsc{p}}}\stackrel{\text{\textsc{n}}\rightarrow\infty}{\longrightarrow}\: & \frac{1}{2}+\frac{1}{2}\int_{-\pi}^{\pi}\frac{dz}{2\pi}\,\frac{(\frac{1}{2}\omega+2J\cos z)^{2}}{(\frac{1}{2}\omega+2J\cos z)^{2}+4J^{2}\sin^{2}z}\nonumber \\
= & \begin{cases}
3/4, & |\tilde{J}|\ge1/2\\
1-\tilde{J}^{2}, & |\tilde{J}|<1/2
\end{cases}.
\end{align}
$\tilde{\text{\textsc{n}}}_{\text{\textsc{p}}}(\tilde{J})$ has a
singular point at $\tilde{J}\equiv J/\omega=1/2$, and this is just
the critical point when discussing the ground state phase transition
of the quantum Ising model.

In this sense, we adopt $\langle\hat{\text{\textsc{n}}}_{\mathrm{t}}\rangle\simeq\tilde{\text{\textsc{n}}}_{\text{\textsc{p}}}$
as an approximated constraint, and estimate the possible maximum of
the total correlation by variation. As mentioned below Eq.\,(\ref{eq:pe}),
the density state $\rho_{n}$ of each site keeps diagonal during the
evolution, thus its entropy is $S[\rho_{n}]=-p_{n,\mathrm{e}}\ln p_{n,\mathrm{e}}-p_{n,\mathrm{g}}\ln p_{n,\mathrm{g}}$
{[}$p_{n,\mathrm{g(e)}}$ is the probability that site-$n$ is in
the ground (excited) state{]}. The full $N$-body state keeps a pure
one, whose entropy is zero. With the help of Lagrangian multipliers,
under the constraints (1) $p_{n,\mathrm{g}}+p_{n,\mathrm{e}}=1$,
(2) $\langle\hat{\text{\textsc{n}}}_{\mathrm{t}}\rangle=\sum_{n}\,p_{n,\mathrm{e}}\simeq\tilde{\text{\textsc{n}}}_{\text{\textsc{p}}}$,
the maximum of the total correlation $\mathbf{C}=\sum_{n}\,-p_{n,\mathrm{e}}\ln p_{n,\mathrm{e}}-p_{n,\mathrm{g}}\ln p_{n,\mathrm{g}}$
is calculated by taking variation of the functional 
\begin{align}
F\big[\{p_{n,\mathrm{e}}\},\lambda\big]:= & \big(\sum_{n}-p_{n,\mathrm{e}}\ln p_{n,\mathrm{e}}-p_{n,\mathrm{g}}\ln p_{n,\mathrm{g}}\big)\nonumber \\
 & +\lambda\big(\tilde{\text{\textsc{n}}}_{\text{\textsc{p}}}-\sum_{n}p_{n,\mathrm{e}}\big)
\end{align}
 upon $p_{n,\mathrm{e}}$ and $\lambda$. 

To look for the extremum, $\partial F/\partial p_{n,\mathrm{e}}\equiv0=\ln(1-p_{n,\mathrm{e}})-\ln p_{n,\mathrm{e}}-\lambda$
indicates all $p_{n,\mathrm{e}}$ should be equal to each other. Thus
the total correlation $\mathbf{C}[\{p_{n,\mathrm{e}}\}]$ achieves
its possible maximum when all the $N$ sites take $p_{n,\mathrm{e}}=\tilde{\text{\textsc{n}}}_{\text{\textsc{p}}}/N$.
As a result, the correlation maximization gives 
\begin{equation}
\mathbf{C}_{\text{max}}^{(N)}=-\tilde{\text{\textsc{n}}}_{\text{\textsc{p}}}\ln\frac{\tilde{\text{\textsc{n}}}_{\text{\textsc{p}}}}{N}-(N-\tilde{\text{\textsc{n}}}_{\text{\textsc{p}}})\ln(1-\frac{\tilde{\text{\textsc{n}}}_{\text{\textsc{p}}}}{N}).\label{eq:C}
\end{equation}

In the weak coupling regime {[}Fig.\,\ref{fig:TCE}(b, c){]} we see
this correlation maximum (the dashed horizontal lines) is quite close
to the time dependent result that the total correlation is increasing
towards. In the strong coupling regime {[}Fig.\,\ref{fig:TCE}(e,
f){]}, though containing heavy oscillations, the envelope of the total
correlation $\mathbf{C}(t)$ also coincide well with the possible
maximum $\mathbf{C}_{\text{max}}^{(N)}$.

\begin{figure}
\includegraphics[width=1\columnwidth]{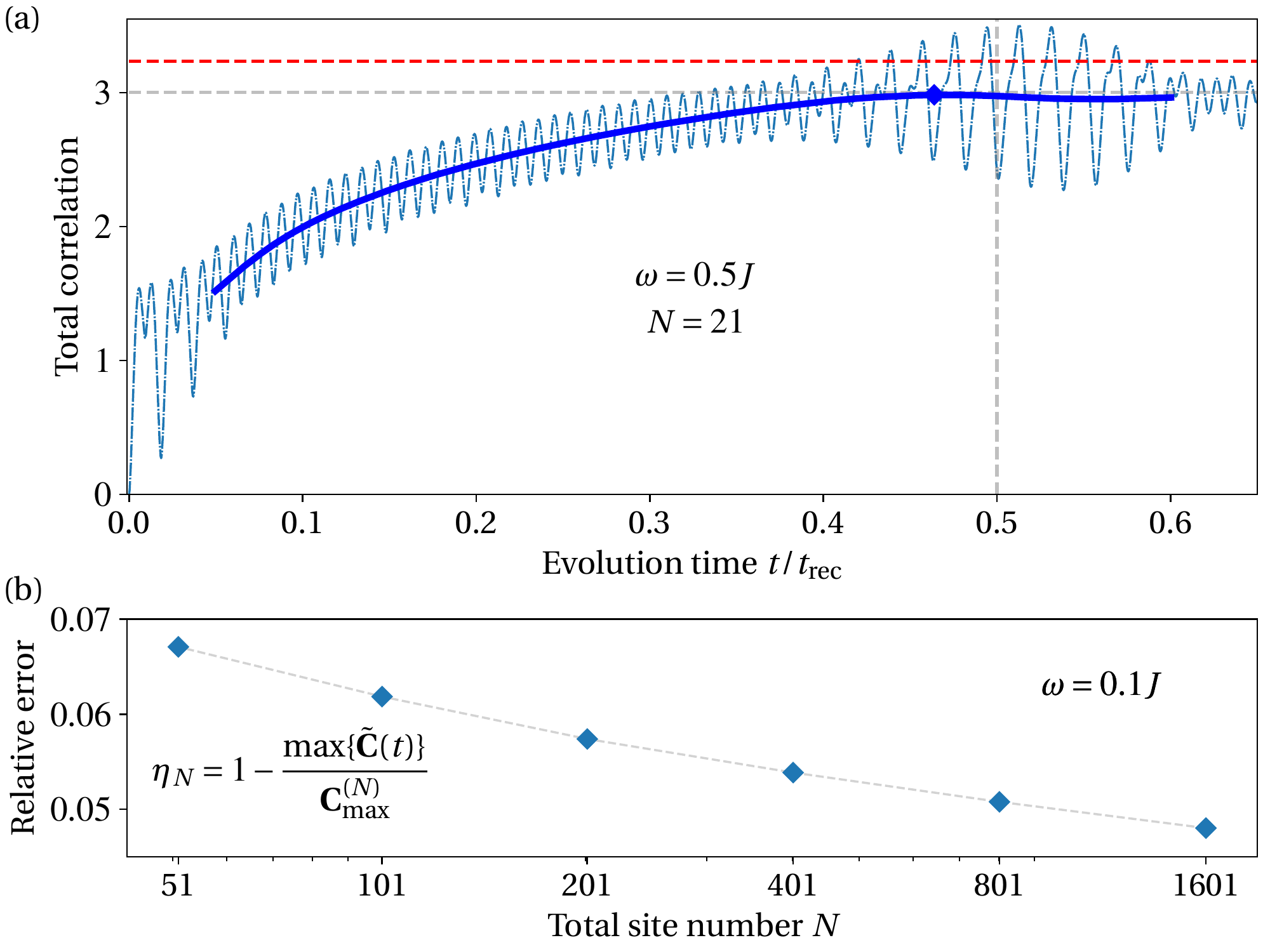}

\caption{(a) Demonstration for the total correlation after coarse-graining
(thick blue line, $\omega=0.1J$, $N=21$). The dash-dotted line is
the original total correlation $\mathbf{C}(t)$ without coarse-graining.
The dashed red line is the possible correlation maximum $\mathbf{C}_{\text{max}}^{(N)}$.
(2) The relative error between $\mathbf{C}_{\text{max}}^{(N)}$ and
the maximum of the coarse-grained correlation $\text{max}\big\{\tilde{\mathbf{C}}(t)\big\}$
around $t\simeq t_{\text{rec}}/2$ for different system sizes ($N=51,\,101,\,201,\,401,\,801,\,1601$,
$\omega=0.1J$).}

\label{fig:error}
\end{figure}

To make a more precise comparison, here we consider the coarse-grained
total correlation, i.e., 
\begin{equation}
\tilde{\mathbf{C}}(t):=\frac{1}{2\Delta t}\int_{t-\Delta t}^{t+\Delta t}ds\,\mathbf{C}(s).
\end{equation}
 Namely, at time $t$, $\tilde{\mathbf{C}}(t)$ is a time average
over a small period $2\Delta t$ in which many enough oscillations
are included \citep{gibbs_elementary_1902,li_correlation_2019}. In
this sense $\tilde{\mathbf{C}}(t)$ gives the central line of the
exact $\mathbf{C}(t)$ {[}the thick blue line in Fig.\,\ref{fig:error}(a){]},
which turns out to be a smoothly increasing line. The maximum of $\tilde{\mathbf{C}}(t)$
appears around $t\simeq t_{\text{rec}}/2$, and we compare this $\text{max}\{\tilde{\mathbf{C}}(t)\}$
with the possible maximum $\mathbf{C}_{\text{max}}^{(N)}$. With the
increase of the system size $N$, the relative error between $\text{max}\{\tilde{\mathbf{C}}(t)\}$
and $\mathbf{C}_{\text{max}}^{(N)}$ decreases {[}see Fig.\,\ref{fig:error}(b){]}.
Thus, when the size $N\rightarrow\infty$, it is expectable that the
coarse-grained total correlation $\tilde{\mathbf{C}}(t)$ is just
increasing towards $\mathbf{C}_{\text{max}}^{(N)}$ as its destination.

Therefore, here the behavior of the total correlation  in this system
well manifests the irreversible entropy production in the standard
macroscopic thermodynamics. Remember that the standard entropy production
based on thermal entropy $dS=\text{\dj}Q/T$ cannot be applied for
the nonequilibrium states here.

\section{Reducing to entropy production in standard thermodynamics \label{sec:REDUCING TO }}

Now we show that, under proper physical conditions, the total correlation
could reduce to the entropy production in the standard thermodynamics. 

In most open system problems, we consider an open system interacting
with a bath, and the bath is usually modeled as a collection of many
noninteracting DoFs $\hat{H}_{\textsc{b}}=\sum_{n}\hat{\textsc{h}}_{n}$,
which is weakly coupled with the open system, and it is assumed that
the initial state of the bath is a canonical thermal state $\hat{\rho}_{\text{\textsc{b}}}(0)\sim\exp(-\hat{H}_{\text{\textsc{b}}}/k_{\text{\textsc{b}}}T)$.
When the system-bath interaction strength is negligibly small, since
the bath is much larger than the open system, the bath state $\hat{\rho}_{\text{\textsc{b}}}(t)$
would not change too much comparing with its initial one, thus $\ln\hat{\rho}_{\text{\textsc{b}}}(t)\simeq\ln\hat{\rho}_{\text{\textsc{b}}}(0)+o[\delta\hat{\rho}_{\text{\textsc{b}}}(t)]$,
and the changing rate of the bath entropy can be approximately obtained
as \citep{li_production_2017,aurell_von_2015,you_entropy_2018,manzano_quantum_2018}
\begin{align}
\dot{S}_{\text{\textsc{b}}}(t) & =-\frac{d}{dt}\mathrm{tr}\left[\hat{\rho}_{\text{\textsc{b}}}(t)\ln\hat{\rho}_{\text{\textsc{b}}}(t)\right]=-\mathrm{tr}\big[\dot{\hat{\rho}}_{\text{\textsc{b}}}(t)\ln\hat{\rho}_{\text{\textsc{b}}}(t)\big]\nonumber \\
 & \simeq-\mathrm{tr}\big[\dot{\hat{\rho}}_{\text{\textsc{b}}}(t)\ln\hat{\rho}_{\text{\textsc{b}}}(0)\big]=\frac{1}{T}\frac{d}{dt}\langle\hat{H}_{\text{\textsc{b}}}\rangle.
\end{align}

Notice that here the bath energy increase $\frac{d}{dt}\langle\hat{H}_{\text{\textsc{b}}}\rangle$
is just equal to the system energy loss $-\dot{Q}$, namely, the informational
entropy change $\dot{S}_{\text{\textsc{b}}}$ could reduce to the
thermal entropy (up to a minus sign), $dS_{\text{\textsc{b}}}=-\text{\dj}Q/T$.
On the other hand, under the weak interactions, the different DoFs
in the bath would not generate significant correlations during the
evolution, thus approximately the bath entropy is just the summation
of the entropy of each DoF, $S_{\text{\textsc{b}}}(t)\simeq\sum_{n}\,S[\hat{\varrho}_{\text{\textsc{b}},n}(t)]$,
where $\hat{\varrho}_{\text{\textsc{b}},n}$ is the reduced density
state of a certain DoF in the bath. Further, the open system plus
the bath as a whole isolated system follows the unitary evolution,
thus the entropy $S_{\text{\textsc{sb}}}$ of the full system does
not change with time. Thus, the changing rate of the total correlation
gives 
\begin{equation}
\frac{d}{dt}\mathbf{C}=\frac{d}{dt}(S_{\text{\textsc{s}}}+S_{\text{\textsc{b}}}-S_{\text{\textsc{sb}}})\simeq\frac{d}{dt}S_{\text{\textsc{s}}}-\frac{1}{T}\frac{dQ}{dt}\ge0,
\end{equation}
which returns the irreversible entropy production in the standard
thermodynamics \citep{li_production_2017,you_entropy_2018,li_correlation_2019}.
In this sense, the second law statement that the irreversible entropy
keeps increasing could be equivalently understood as the increase
of total correlation in this whole system \citep{esposito_entropy_2010,li_production_2017,manzano_entropy_2016,alipour_correlations_2016,pucci_entropy_2013,hobson_irreversibility_1966,zhang_general_2008,horowitz_equivalent_2014,kalogeropoulos_time_2018}.
If the correlation between the different DoF inside the bath is significant,
corrections should be considered \citep{ptaszynski_entropy_2019}.

Another analogy with the standard thermodynamics is the Boltzmann
entropy increase in an isolated classical gas composed of $N$ particles
with weak collisions. Since the full ensemble state $\rho(\vec{P},\vec{Q})$
of the \emph{N}-body system follows the Liouville theorem, its Gibbs
entropy \citep{lebowitz_macroscopic_1993,li_correlation_2019},

\begin{equation}
S_{\textsc{g}}[\rho(\vec{P},\vec{Q})]=-\int d^{3N}pd^{3N}q\,\rho\ln\rho,
\end{equation}
never changes with time. One the other hand, the dynamics of the single
particle probability distribution function (PDF) $\varrho_{n}(\mathbf{p}_{n},\mathbf{x}_{n})$
can be described by the Boltzmann equation. According to the Boltzmann
\emph{H} theorem \citep{landau_statistical_1980}, because of the
particle collisions the PDF $\varrho_{n}(\mathbf{p}_{n},\mathbf{x}_{n})$
of each single particle always approaches the Maxwell-Boltzmann distribution
($\sim\exp[-E(\mathbf{p}_{n},\mathbf{x}_{n})/k_{\text{\textsc{b}}}T]$)
as the steady state, with its entropy increasing monotonically. Notice
that the single-particle PDF $\varrho_{n}(\mathbf{p}_{n},\mathbf{x}_{n})$
is a marginal distribution of the full ensemble state $\rho(\vec{P},\vec{Q})$,
which is obtained by averaging out all the other particles. In this
case the total correlation  of the $N$ particle gas is 
\begin{equation}
\mathbf{C}=\sum_{n}S_{\textsc{g}}[\varrho_{n}(\mathbf{p}_{n},\mathbf{x}_{n})]-S_{\textsc{g}}[\rho(\vec{P},\vec{Q})],
\end{equation}
which sums up the entropy of all the $N$ single particle PDF, and
$S_{\textsc{g}}[\rho(\vec{P},\vec{Q})]$ always keeps a constant during
the evolution. Therefore, it turns out the increase of the total correlation
$\delta\mathbf{C}(t)$ is just equal to the total Boltzmann entropy
increase of the $N$ particle gas. Therefore, the total correlation
 reduces to the entropy production in the standard thermodynamics
\citep{lebowitz_macroscopic_1993,li_correlation_2019,jaynes_gibbs_1965,chliamovitch_kinetic_2017}.

In practice, usually it is the partial information (e.g., marginal
distribution, few-body observable expectations) that is directly accessible
to our observation. Indeed most macroscopic thermodynamic quantities
are obtained only from such partial information like the one-body
distribution, which exhibits irreversible behaviors. In practical
measurements, the dynamics of the full many body state is quite difficult
to be measured directly. In this sense, the reversibility of microscopic
dynamics and the macroscopic irreversibility coincide with each other.

\section{Conclusion \label{sec:CONCLUSION}}

Here we consider the nonequilibrium dynamics in an isolated $N$-body
system with Ising interaction. During the unitary evolution, without
applying time average, the full $N$-body state always keeps a constant
entropy, and the entropy of each single site increases and decreases
from time to time. In comparison, the dynamics of the local observables
exhibit local relaxation behavior, which is similar as the macroscopic
thermodynamics. Due to the finite system size, recurrences appears
in the local relaxations which roots from the superposition with the
propagation regathered back. The total correlation  approximately
exhibits a monotonic increasing behavior, which is similar to the
irreversible entropy increase in the standard thermodynamics. Even
for the strong coupling situation where the total excitation has violent
fluctuations, the total excitation exhibits an explicit converging
behavior until the recurrence happens.We also show that in most common
physical conditions, the correlational entropy production could reduce
to the standard entropy production based on the thermal entropy. In
this sense, the correlational entropy production is a well generalization
for the standard entropy production.

\end{document}